\documentclass[10pt,a4paper,oneside,onecolumn]{article}
\usepackage[T1]{fontenc}
\usepackage[left=2cm, right=2cm, top=2cm, bottom=2cm]{geometry}

\usepackage{graphicx}

\usepackage{indentfirst,csquotes}

\usepackage{xcolor}
\usepackage{balance}

\usepackage{cite}
\usepackage{amsmath,amssymb,amsfonts}
\usepackage{algorithmic}
\usepackage{graphicx}
\usepackage{textcomp}
\usepackage{array}
\usepackage{multirow}
\usepackage{balance}

\usepackage{steinmetz}

 \usepackage[caption=false,font=normalsize,labelfont=sf,textfont=sf]{subfig}

\begin{document}

\title{Low-profile Super-Realised Gain Antennas}

\author{
James Moore$^{1}$,
Aaron M. Graham$^{1}$,
Manos M. Tentzeris$^{2}$,
Vincent Fusco$^{1}$,\\
Stylianos D. Asimonis$^{1}$
\\
\small {$^{1}$Electronics Communications and Information Technology, ECIT, Queen’s University Belfast, Belfast, U.K.}
\\
\small {$^{2}$Georgia Institute of Technology, School of Electrical and Computer Engineering, Atlanta, GA 30332, USA
}
\\
\small {rmoore57@qub.ac.uk, agraham63@qub.ac.uk, etentze@ece.gatech.edu, v.fusco@ecit.qub.ac.uk, s.asimonis@qub.ac.uk
}
}

\date{} 

\maketitle


\begin{abstract}
This study introduces an novel approach to the design of low-profile superdirective antenna arrays, employing parasitic elements. The proposed design concept was verified through the analysis of a two-element antenna array consisting of strip dipoles, operating at a frequency of $3.5$ GHz within the sub-6 5G frequency band. The antenna array was optimized for realized gain, a key antenna parameter considering both ohmic and return losses. The design parameters included the lengths and widths of the strip dipoles, along with a reactance load connected to the parasitic element. The driven element was excited by a sinusoidal voltage signal with a magnitude of $1$ V, eliminating the need for amplifiers, attenuators, phase shifters, or impedance matching networks. Results demonstrated that this design concept enables the achievement of superdirectivity, with inter-element distances as small as $0.1\lambda$, resulting in low-profile, compact, high directional antenna systems.

\vskip0.5\baselineskip
\end{abstract} 

\bigskip

\section{Introduction}

In general, precise and focused signal transmission and reception improves communication performance over long distances. Therefore, the design of directional antennas is important in wireless communications. In \cite{uzkov1946approach } Uzkov  proved theoretically that in a uniform linear array (ULA) of $N$ isotropic radiators, the directivity is end-fire and tends to $N^2$ as the inter-element distance tends to zero and as the elements are driven by properly chosen voltage signals. This phenomenon, known as \textit{superdirectivity}, pointed the way to the design of highly directional antenna arrays \cite{Yaru1951,harrington1958gain,altshuler2005monopole,Morris2005Superdirectivity,Kim2012Superdirective,dovelos2022superdirective,Marzetta2019Super,dovelos2023MIMO,Assimonis2023How}. In \cite{hansen1981fundamental}, Hansen proposed a criterion for characterizing an antenna array as superdirective; he stated that an antenna array can be considered a superdirective if its directivity is higher than that of an identical antenna array uniformly excited over all elements. However, the very close placement of the elements leads to strong mutual electromagnetic coupling between the elements of the array, which in turn leads to high corresponding input currents. Thus, superdirective antenna arrays suffer from high intrinsic ohmic losses, resulting in low radiation efficiency and gain. Most importantly, the input impedance of each radiating element exhibits high reactance due to this strong coupling. The latter complicates efforts to achieve conventional $50, \Omega$ impedance matching, resulting in to low achieved realized gain. Therefore, the implementation of superdirective antenna arrays necessitates the incorporation of amplifiers and phase shifters to adjust the excitation voltage signals, along with external impedance matching networks. These additional components significantly augment the complexity of these designs.

To address these challenges, researchers in \cite{Lynch2023} have introduced a novel approach to designing superdirective antennas. They form antenna arrays of strip dipole elements of slightly different lengths and widths, excited by signals of equal magnitude but different phase. This configuration eliminates the need for amplifiers. The appropriate phase difference between the excitation signals was achieved by incorporating waveguides (specifically coaxial cables) of different lengths. Moreover, the researchers in \cite{Lynch2023} emphasized the importance of considering ohmic and return losses. Thus, they optimized the design to maximize the realized gain. However, the need for accurate phase difference adjustment between the radiating elements keeps the design complexity high.
This paper presents a novel approach to the design of superdirecive antenna arrays where only a single element is driven while the others are loaded. This method leads to superdirectional antenna systems with high radiation efficiency and low return losses, i.e. with high realized gain, without the use of amplifiers, phase shifters and external impedance matching networks.

\newpage
\section{Antenna Design}

\begin{figure}[h!]
 \centering
 \includegraphics[width=0.5\linewidth]{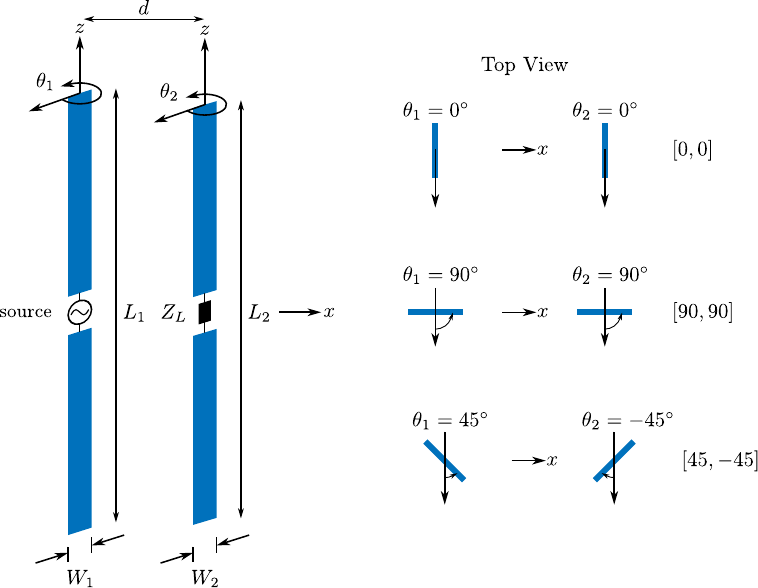}
 \caption{ \footnotesize {Configuration of the microstrip dipole loaded array, with elements of lengths \emph{L$_1$} and \emph{L$_2$}, widths \emph{W$_1$} and \emph{W$_2$}, inter-element distance \emph{d} and the tilt of each dipole $\theta_1$ and $\theta_2$.}}
 \label{fig:Geometry}
 \end{figure}

The superdirective antenna array was designed to operate at $3.5$ GHz, targeting the fifth generation (5G) sub-6 (below 6 GHz) frequency band. Strip dipoles were chosen as radiating elements due to their simplicity. The analyzed antenna array is depicted in Fig. \ref{fig:Geometry}; it consists of two strip dipole elements with lengths $L_1$ and $L_2$, widths $W_1$ and $W_2$, placed at a distance $d$ along the $x$-axis. The elements can be tilted by angles $\theta_1$ for the first element and $\theta_2$ for the second element, around the $z$-axis. Full electromagnetic numerical simulations were conducted using the Antenna Toolbox of MATLAB \cite{matlab2022}, with the Method of Moments applied. To ensure accuracy, the maximum edge length for the mesh was set to $\lambda/41$, where $\lambda$ is the wavelength at $3.5$ GHz in free space. The conductor strips were made of copper, with a conductivity of $5.8\times 10^7$ S/m and a thickness of $35$ $\mu$m. The first element was driven by a fixed voltage sinusoidal signal with a magnitude of $1$ V, while the second element was not excited, but a load $Z_L$ was connected in the middle of the strip dipole, between its two arms. Thus, the proposed technique will henceforth be referred to as the \textit{loaded} case. In fact, the second element acts as a parasitic element.

Three orientations of the strip dipoles were considered in our calculations: first, the dipoles were fixed at $\theta_1=\theta_2=0$ (case $[0, 0]$), where both flat sides of the dipoles faced each other; next, $\theta_1=45^{\circ}$, $\theta_2=-45^{\circ}$ (case $[45, -45]$) where the flat sides were perpendicular to each other; and finally, $\theta_1=\theta_2=90^{\circ}$ (case $[90, 90]$) where the edges of the dipoles faced each other.
The antenna array was optimized to achieve maximum realized gain in the horizontal plane (i.e., $xy$-plane), considering the strip dimensions $L_i$ and $W_i$, where $i=1,2$, and the load $Z_L$. Please note that the connected load is assumed to be purely reactive, with zero resistance, in order to avoid additional ohmic losses. Particle Swarm Optimization (PSO) was chosen for its ability to efficiently explore complex solution spaces and find global optima. It is versatile, easy to implement, robust against local minima, does not require gradient information, and has few tunable parameters, making it well-suited for antenna design and similar optimization tasks \cite{ParticleSwarm}. Mathematically, the optimization problem is expressed as:

\begin{equation}
\begin{aligned}
\underset{
\left\lbrace L_1,L_2,W_1,W_2,Z_L \right\rbrace 
}{\text{Maximize}}
& \quad f \left(L_1,L_2,W_1,W_2,Z_L\right) \\
\text{subj. to:} 
        & \quad L_1, L_2 \in \left[0.3\lambda, 0.7\lambda\right],\\
        & \quad W_1, W_2 \in \left[10^{-3}\lambda, 10^{-1}\lambda\right],\\
        & \quad Z_L \in \left[-j 1~\mathrm{k}\Omega, j 4~\mathrm{k}\Omega\right],
\end{aligned}
\end{equation}
where $f$ denotes the realized gain function.
\begin{figure*}[b!]
    \centering
    \subfloat[]{\includegraphics[width=0.33\columnwidth]{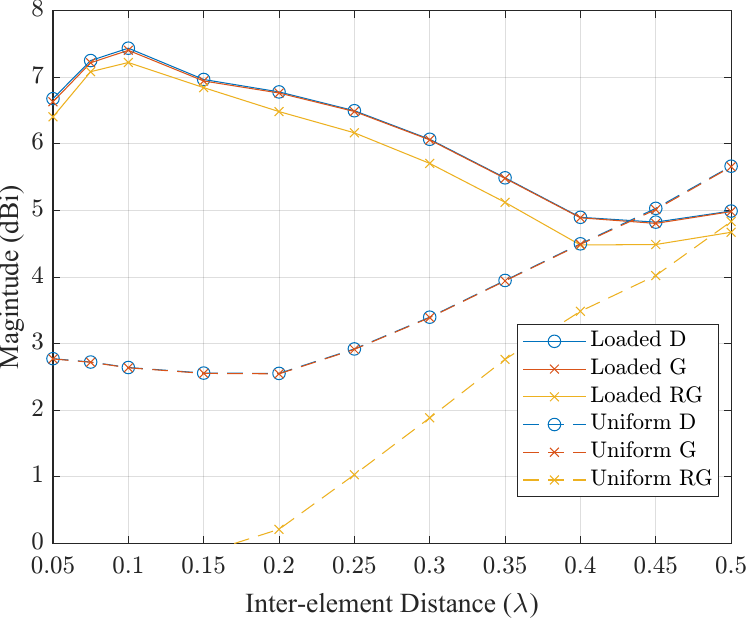}}
    \hfil
    \subfloat[]{\includegraphics[width=0.33\columnwidth]{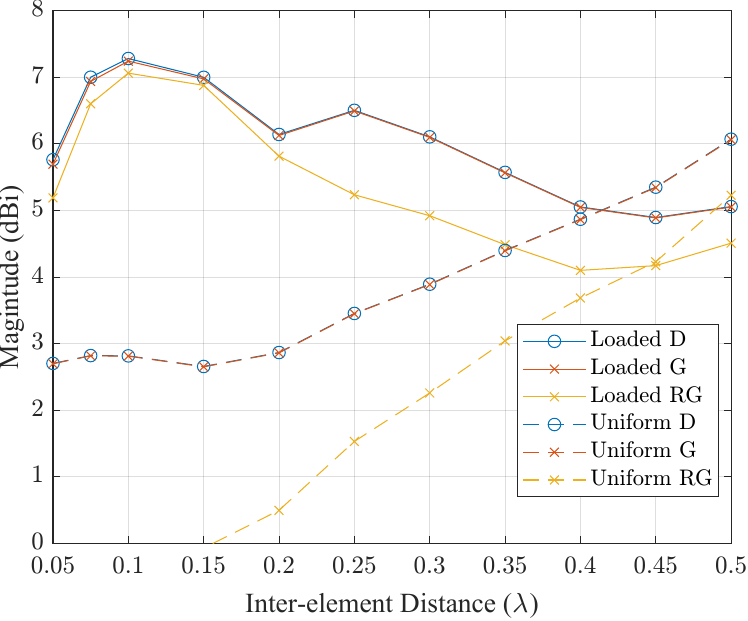}}
    \hfil
    \subfloat[]{\includegraphics[width=0.33\columnwidth]{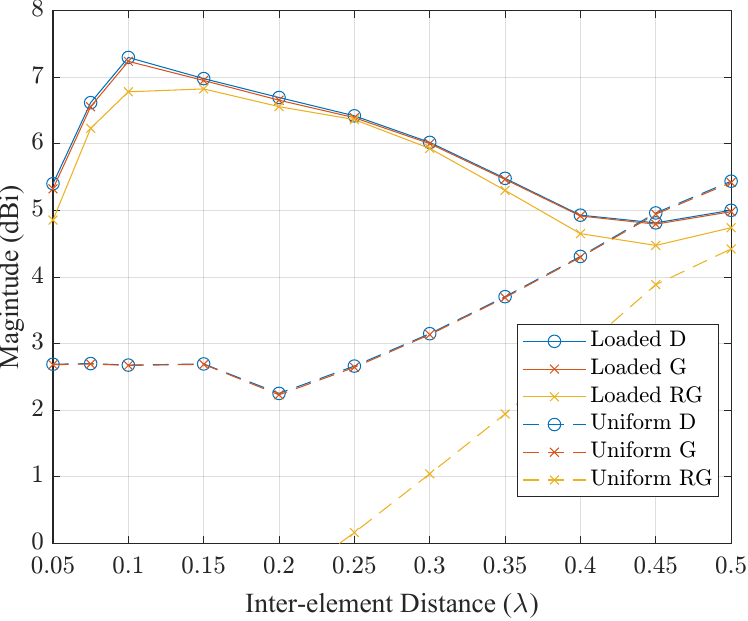}}
    \hfil
    \subfloat[]{\includegraphics[width=0.33\columnwidth]{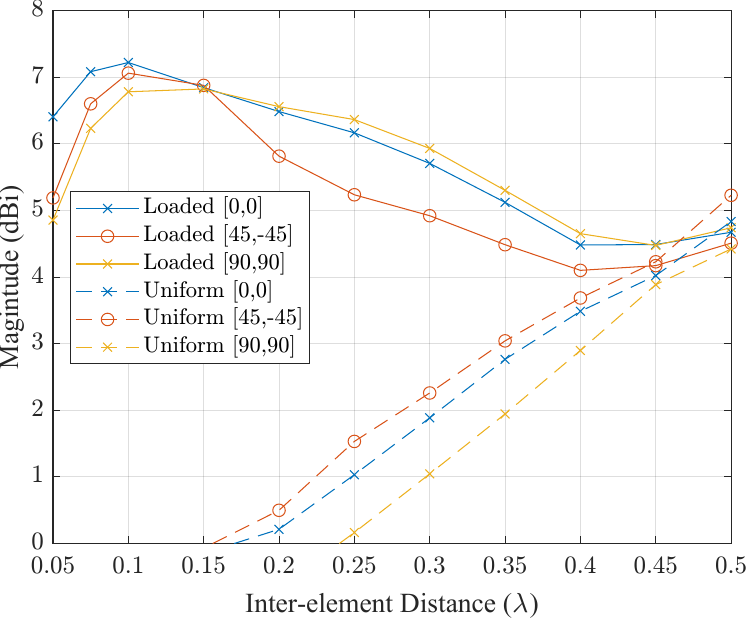}}
    \hfil
    \subfloat[]{\includegraphics[width=0.33\columnwidth]{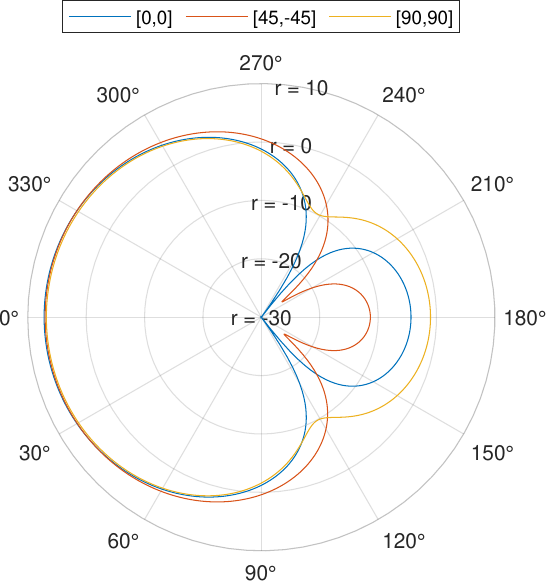}}
    \hfil
    \subfloat[]{\includegraphics[width=0.33\columnwidth]{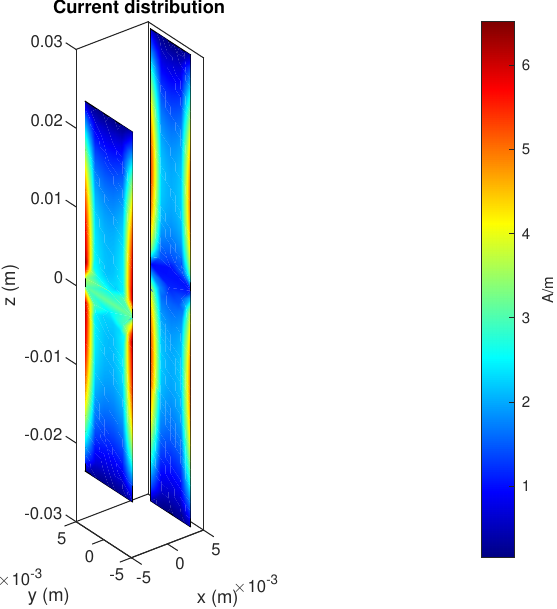}}
    \caption{
   \footnotesize { Overall maximum directivity, gain, and realized gain in the horizontal plane against the inter-element distance for three orientations: $[0, 0]$ (a), $[45, -45]$ (b), and $[90, 90]$ (c), comparing both the loaded antenna array and the uniformly excited equivalent array. Overall maximum realized gain for each orientation versus inter-element distance, comparing the loaded antenna array to the uniformly excited equivalent array (d). Realized gain of the loaded antenna array in the horizontal plane at each orientation (e). Current distribution for the $[0, 0]$ orientation (f).}}
    \label{fig:OrientationResults}
\end{figure*}

The realized gain was optimized for each of the three orientations at different inter-element distances $d$. 
For the optimal design parameters, the total maximum realized gain in the horizontal plane is depicted in Fig. \ref{fig:OrientationResults}(a) to Fig. \ref{fig:OrientationResults}(c). The corresponding directivity and gain are also shown for the sake of clarity. In order to validate the criterion of superdirectivity performance of the proposed antenna array based on Hansen's work \cite{hansen1981fundamental}, directivity, gain and realized gain of an array with exactly the same dimensions but whose elements are uniformly excited (both elements) with sinusoidal voltages of $1$ V are also shown in the same graph.
It is clear that when the inter-element distances are small, the antenna exhibits superdirectional performance, i.e., the maximum realized gain of the loaded case is significantly higher than that of the uniform excitation case for all orientations. 
\newpage
This trend is also observed when comparing directivity and gain between the loaded and the uniform cases. Specifically, for all orientations the maximum realized gain occurs at  $d=0.1\lambda$. However, beyond about $0.4\lambda$, the proposed technique fails to achieve a realized gain higher than that of the uniform excitation case. This result was expected since the super-directional phenomenon occurs mainly at small inter-element distances. Fig. \ref{fig:OrientationResults}(d) compares only the realized gain for each orientation (and the uniform counterpart). It is clear that the $[0, 0]$ orientation performs best for very small interelement distances $\leq 0.15 \lambda$, in addition to having the highest maximum realized gain. In contrast, $[90, 90]$ produces the lower maximum realized gain, but performs best at larger distances $\geq 0.2 \lambda$. The close agreement between directivity, gain, and realized gain for $[90, 90]$ indicates that it is much easier to maintain a good impedance matching at larger inter-element distances than for the other orientations. In this benchmark, the $[45, -45]$ is undoubtedly the poorest performer. However, Fig. \ref{fig:OrientationResults}(e) paints a different picture, showing the radiation pattern in the horizontal plane. As can be seen, the radiation pattern is unique for each orientation. Although the main lobes of each orientation are roughly equivalent in maximum and beamwidth, the $[45, -45]$ configuration produces the smallest front to back ratio (F/B ratio). With an F/B ratio of $18$ dB, $[45, -45]$ outperforms the other two orientations with $[0, 0] $ at $12.3$ dB and $[90, 90] $ at $8$ dB. Thus, if an F/B ratio is desired, then the $[45, -45]$ configuration is the most desirable in this regard. However, $[0, 0]$ orientation has the best overall performance, with the highest maximum realized gain over a wide range of inter-element distances, as well as a F/B ratio greater than $10$ dB. 

For the $[0, 0]$ orientation, the current distribution is shown in Fig. \ref{fig:OrientationResults}(f). It can be seen that the current is heavily concentrated along the vertical edges of the dipoles, evenly distributed on both sides in symmetry. Therefore, it can be concluded that the orientations $[0, 0]$ and $[90, 90]$ are more sensitive to mutual coupling than the $[45, -45]$ case. This is because the edges of the strips are more symmetrically adjacent in these orientations compared to the $[45, -45]$ case. Additionally, as explained earlier, superdirectivity necessitates strong coupling. Hence, this mechanism clarifies why the maximum realized gain decreases more rapidly in the $[45, -45]$ case than in the $[0, 0]$ or $[90, 90]$ cases as the inter-element distance increases.

\newpage
Table \ref{DimTable} lists the optimal dimensions and load impedance from $0.05 \lambda$ to $0.5 \lambda$. The maximum directivity, gain, and realized gain of the main lobe in the horizontal plane of these maximums are also listed. In order to implement the loaded element in practice, the load impedance can be realized through inserting equivalent capacitors for a negative reactance value, i.e., $Z_L = -j(2 \pi f C)^{-1}$, and inductors for a positive reactance value, i.e., $Z_L = j 2 \pi f L$, where $C$ and $L$ are the equivalent capacitance and inductance, respectively, and $f$ is the operating frequency. For the values in Table \ref{DimTable}, the following capacitances and inductances can be calculated: $Z_L=-j 1000 \xrightarrow{} C = 45.5$ fF, $Z_L=-j 990 \xrightarrow{} 45.9$ fF, $Z_L = -j 480 \xrightarrow{} 94.7$ fF, and $Z_L = j 4000 \xrightarrow{} 181.4$ nH.

\begin{table*}[h!]
\renewcommand{\arraystretch}{1.2}
    \centering
    \caption{Optimal Design Parameters for the $[0, 0]$ Orientation }
    \label{DimTable}
    \begin{tabular}{
    c|
    c|
    c|
    c|
    c|
    c|
    >{\centering\arraybackslash}m{1.75cm} |
    >{\centering\arraybackslash}m{1.25cm} |
    >{\centering\arraybackslash}m{1.5cm} |
    >{\centering\arraybackslash}m{1.5cm} 
    }
        \hline
        \hline
        $d$ ($\lambda$) & $L_1$ ($\lambda$) & $L_2$ ($\lambda$) & $W_1$ ($\lambda$) & $W_2$ ($\lambda$) & $Z_L$ (k$\Omega$)  & Directivity (dBi) & Gain (dBi) & Realised Gain (dBi) & Direction (deg.) \\
        \hline
        $0.05$ & $0.682$ & $0.698$ & $0.089$ & $0.087$ & $-0.99$ & $6.7$ & $6.6$ & $6.4$ & $0$ \\
        $0.1$ & $0.548$ & $0.698$ & $0.100$ & $0.086$ & $4$ & $7.4$ & $7.4$ & $7.2$ & $0$ \\
        $0.2$ & $0.451$ & $0.698$ & $0.089$ & $0.087$ & $-1$ & $6.8$ & $6.8$ & $6.5$ & $0$ \\
        $0.3$ & $0.451$ & $0.639$ & $0.089$ & $0.089$ & $-1$ & $6.1$ & $6.1$ & $5.7$ & $0$ \\
        $0.4$ & $0.450$ & $0.630$ & $0.089$ & $0.088$ & $-1$ & $4.9$ & $4.9$ & $4.5$ & $39^{\circ}$ \\
        $0.5$ & $0.463$ & $0.698$ & $0.016$ & $0.087$ & $-0.48$ & $5.0$ & $5.0$ & $4.7$ & $75^{\circ}$ \\
        \hline
        \hline
    \end{tabular}
\end{table*}

\begin{figure}[h!]
 \centering
 \includegraphics[width=0.33\linewidth]{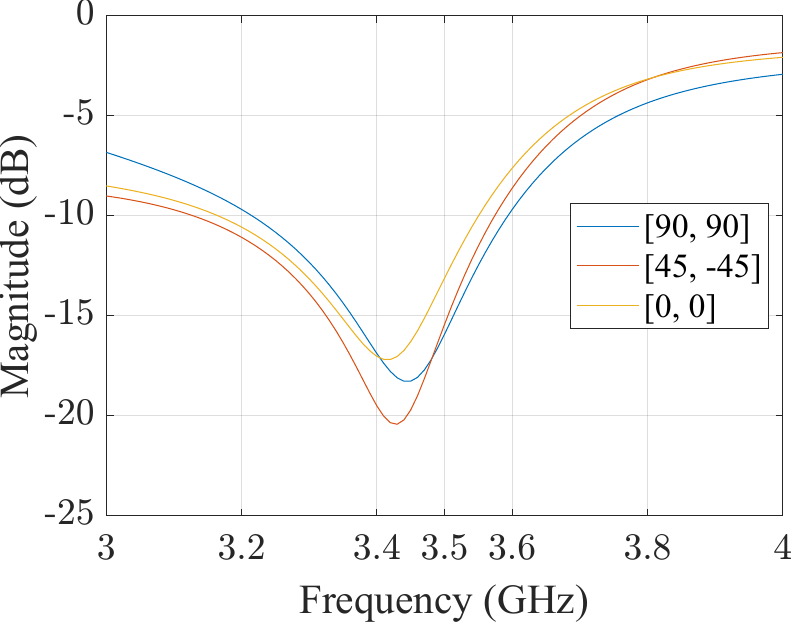}
 \caption{\footnotesize {Reflection coefficient at the driven dipole of the superdirective array, for each orientation.}}
 \label{fig:RefCoef}
 \end{figure}

To evaluate the operating bandwidth of the loaded array, we calculated the reflection coefficient $S_{11}$ for each configuration, as depicted in Fig. \ref{fig:RefCoef}. It is clear that the $[45 ,-45]$  orientation had the widest bandwidth, ranging from $3.13$ GHz to $3.58$ GHz, resulting in a fractional bandwidth (FBW) of $13.4\%$. Close behind, the $[90, 90]$ tilt offered a bandwidth of $3.2$ GHz to $3.6$ GHz (i.e., $\mathrm{FBW}=11.8\%$). Finally, the [$0, 0$] orientation had the narrowest bandwidth, ranging from $3.18$ GHz to $3.55$ GHz (i.e., $\mathrm{FBW}=11\%$).

\section{Conclusion}
In this work, we have demonstrated that the design of a superdirective antenna array can be significantly simplified by incorporating a loaded (or equivalently, a parasitic) radiating element alongside a single driven radiating element, thereby eliminating the need for additional complexities such as attenuators, phase shifters, and amplifiers. Furthermore, our analysis has been focused on the realized gain of the antenna array, taking into account both the ohmic and return losses. The radiating elements utilized were strip dipoles for the sake of simplicity, with the design parameters being their lengths (approximately half-wavelength), widths, and the load connected to the parasitic dipole. The driven dipole was excited by a simple sinusoidal voltage signal. This approach enables the development of low-profile superdirective antenna arrays.

\section*{Acknowledgment}
This work was funded by the Department of the Economy (DfE) in Northern Ireland. As part of their Postgraduate Studentships programme with Queen's University Belfast.

\bibliographystyle{IEEEtran}
\bibliography{mybib}

\end{document}